\input{aipcheck}

\documentclass[
    ,final            
  ]
  {aipproc}

\layoutstyle{6x9}

\begin{document}

\title{Cosmological X-Ray Flashes from Off-Axis Jets}

\author{Ryo Yamazaki}{
  address={Department of Physics, 
           Kyoto University, Kyoto 606-8502, Japan}
}

\author{Kunihito Ioka}{
  address={Department of Earth and Space Science, Osaka University, 
           Toyonaka 560-0043, Japan}
}

\author{Takashi Nakamura}{
  address={Department of Physics, 
           Kyoto University, Kyoto 606-8502, Japan}
}

\newcommand{\bm}[1]{\mbox{{\boldmath $#1$}}} 
\def\lesssim{{\leq}}
\def\gtrsim{{\geq}}
\def\degr{{^\circ}}
\def\N{\nonumber}
\def\f{\frac}
\def\L{\label}
\def\sun{\odot}
\def\VVM{\langle V/V_{\rm max}\rangle}

\begin{abstract}
The $\VVM$ of the cosmological 
X-ray flashes  detected by WFC/{\it BeppoSAX}
is calculated theoretically in  a simple jet model.
The total emission energy from the jet is assumed to be constant.
We find that if the jet opening half-angle is smaller 
than 0.03 radian,
off-axis emission from sources at $z\lesssim4$ can be seen. 
The theoretical $\VVM$ is less than 0.4,
which is consistent with the observational result of
$0.27\pm 0.16$ at the 1\,$\sigma$ level.
This suggests  that the off-axis GRB jet with the  small
opening half-angle at the cosmological distance can be identified
as the cosmological X-ray flash.
\end{abstract}

\maketitle


\section{INTRODUCTION}
The X-ray flash (XRF) is a class of X-ray transients,
whose peak energy of $\nu F_\nu$ spectra is small but
the other properties are roughly similar to those of GRBs
\citep{He01a,ka03,ki02}.
The observational value of $\VVM$ has been updated
from $0.56\pm0.12$ \citep{He00} 
to $0.27\pm0.16$ \citep{He02}.
The updated value of $\VVM$ suggests that XRFs take place at
a cosmological distance.
Various models accounting for the nature of the XRFs
have been proposed
(see \citep{yin03b} and references therein;
see also \citep{lamb03,moch03}).
In our {\it off-axis jet model},
if we observe the GRB jet with a large viewing angle,
it looks like an XRF \citep{yin02,yin03b}.
In \citet{yin02},  the value of 
the jet opening half-angle was  adopted as  $\Delta\theta=0.1$.
Then  the distance to the
farthest XRF ever detected is about 2 Gpc ($z\sim0.4$) 
 so that the cosmological effect is small and 
$\VVM\sim 0.5$.
Recent observations suggest that GRBs with relatively 
small opening angle exist, while 
the distribution of $\Delta\theta$ is not yet clear \citep{pk02}.
If we assume the total emission energy to be constant,
the intrinsic luminosity is larger for the smaller $\Delta\theta$
Such GRBs at the cosmological distance 
observed from off-axis viewing angle 
may be seen as XRFs 
and $\VVM$ is expected to be  smaller than 0.5.

In this paper, we will show that our off-axis model has a
possibility of accounting for the observational value of $\VVM$
if we change some of the model parameters used in 
\citet{yin02}.
%

\section{CALCULATION OF $\protect\VVM$}

We consider a simple jet model of XRFs \citep{yin02,yin03b}
taking into account  the cosmological effect.
The uniform jet with sharp edges is assumed.
See \citet{yin03b} for details.
In order to study the dependence on the viewing angle 
$\theta_v$ and the jet opening half-angle $\Delta\theta$,
we fix the other parameters as
$\alpha_B=-1$, $\beta_B=-3$, 
$\gamma\nu'_0=200\,{\rm keV}$, $r_0/c\beta \gamma^2=10\,{\rm s}$,
and $\gamma=100$.
We fix the amplitude $A_0$ so that
the isotropic $\gamma$-ray energy 
$E_{\rm iso}=4\pi d_L^2 (1+z)^{-1}S_\gamma$ satisfies
$(\Delta\theta)^2E_{\rm iso}/2=0.5\times10^{51}$\,ergs,
when $\theta_v=0$ and $z=1$ \citep{fra01}.
The values of $A_0$ for different opening angles are
summarized in Table~\ref{TableVVM}.
When the jet opening half-angle $\Delta\theta$ becomes smaller,
$A_0$ becomes larger. 

\begin{table}
\begin{tabular}{cccccc}
\hline
   \tablehead{1}{r}{b}{$\Delta\theta$}
  & \tablehead{1}{r}{b}{$A_0$\tablenote{In units of 
                        ${\rm erg}\ {\rm cm}^{-2}\ {\rm Hz}^{-1}$}}
  & \tablehead{1}{r}{b}{$\theta_{v,\, p}$\tablenote{The viewing angle 
                         where $W(\theta_v)$ becomes maximum.}}
  & \tablehead{1}{r}{b}{$z_{\rm max}(\theta_{v,\, p})$}   
  & \tablehead{1}{r}{b}{$z_{\rm min}(\theta_{v,\, p})$}  
  & \tablehead{1}{r}{b}{$\VVM_{\Delta\theta}$\tablenote{
                        For the XRFs detected by WFCs on BeppoSAX.}}
\\
\hline
0.10 & 0.84 & 0.103 & 2.8 & 1.5 & 0.46 \\
0.09 & 1.0 & 0.095 & 2.9 & 1.4 & 0.45 \\
0.08 & 1.3 & 0.086 & 3.0 & 1.4 & 0.44 \\
0.07 & 1.7 & 0.077 & 3.1 & 1.3 & 0.44 \\
0.06 & 2.3 & 0.068 & 3.3 & 1.2 & 0.44 \\
0.05 & 3.4 & 0.060 & 3.5 & 1.2 & 0.44 \\
0.04 & 5.2 & 0.052 & 3.6 & 1.1 & 0.43 \\
0.03 & 9.3 & 0.045 & 3.8 & 0.99  & 0.40 \\
0.02 & 22 & 0.038 & 4.0 & 0.89 & 0.38 \\
0.01 & 109 & 0.034 & 4.1 & 0.77 & 0.35 \\
\hline
\end{tabular}
\caption{
Results of calculation for fixed $\Delta\theta$
}
\label{TableVVM}
\end{table}

\begin{figure}
\includegraphics[width=0.3\textwidth,height=.16\textheight]
                {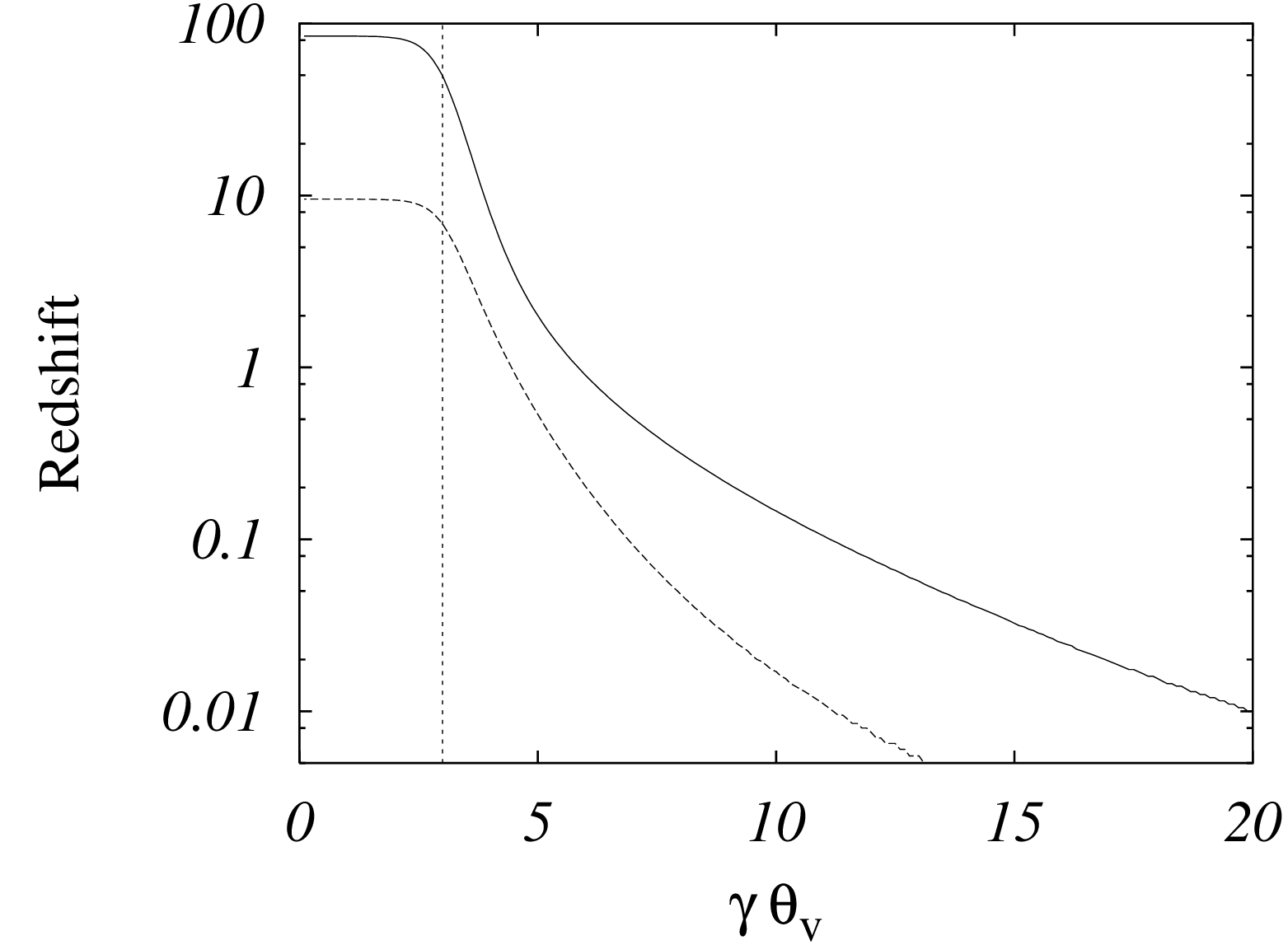}
\hspace{4mm}
\includegraphics[width=0.3\textwidth,height=.16\textheight]
                {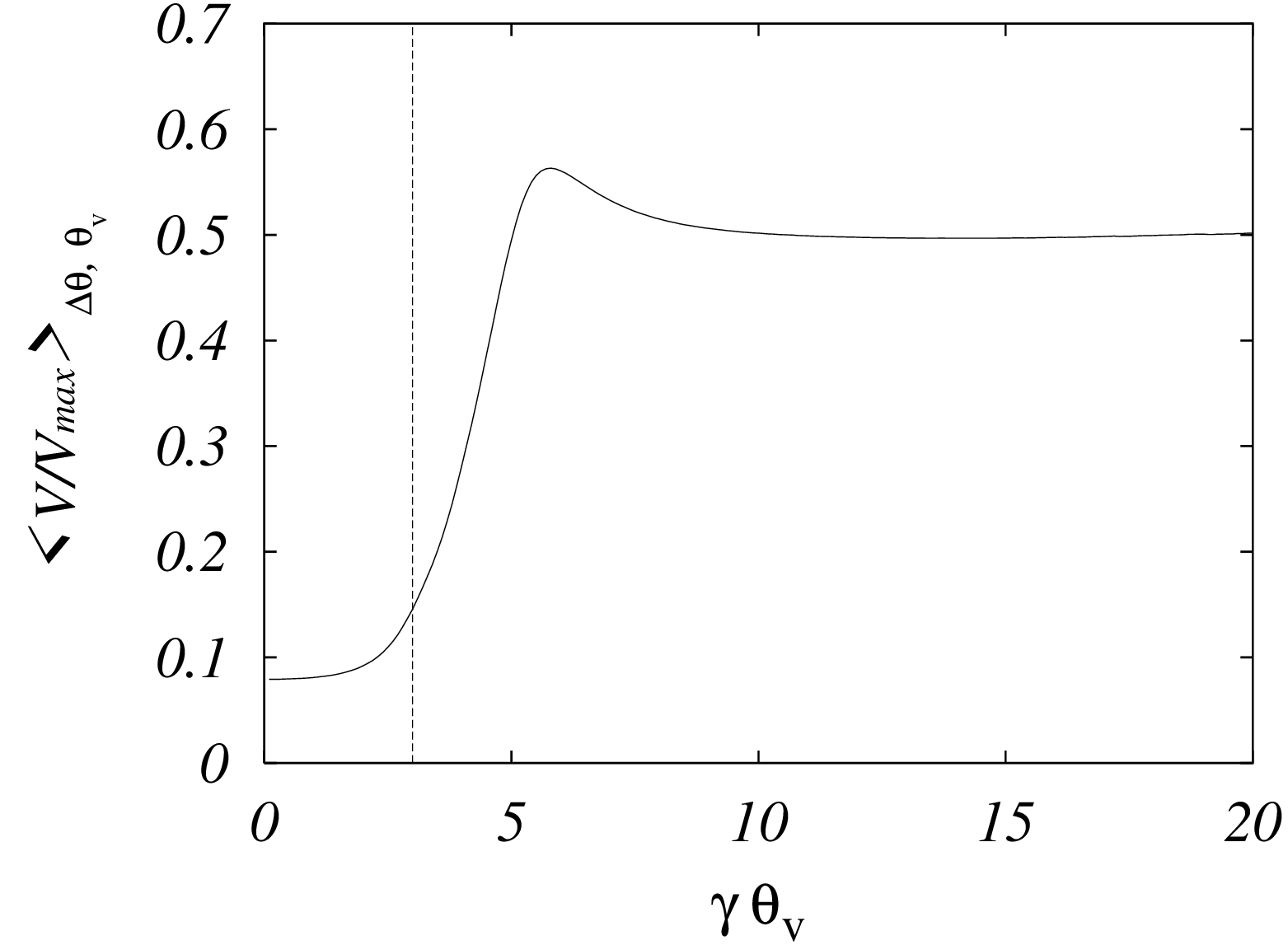}
\hspace{4mm}
\includegraphics[width=0.3\textwidth,height=.16\textheight]
                {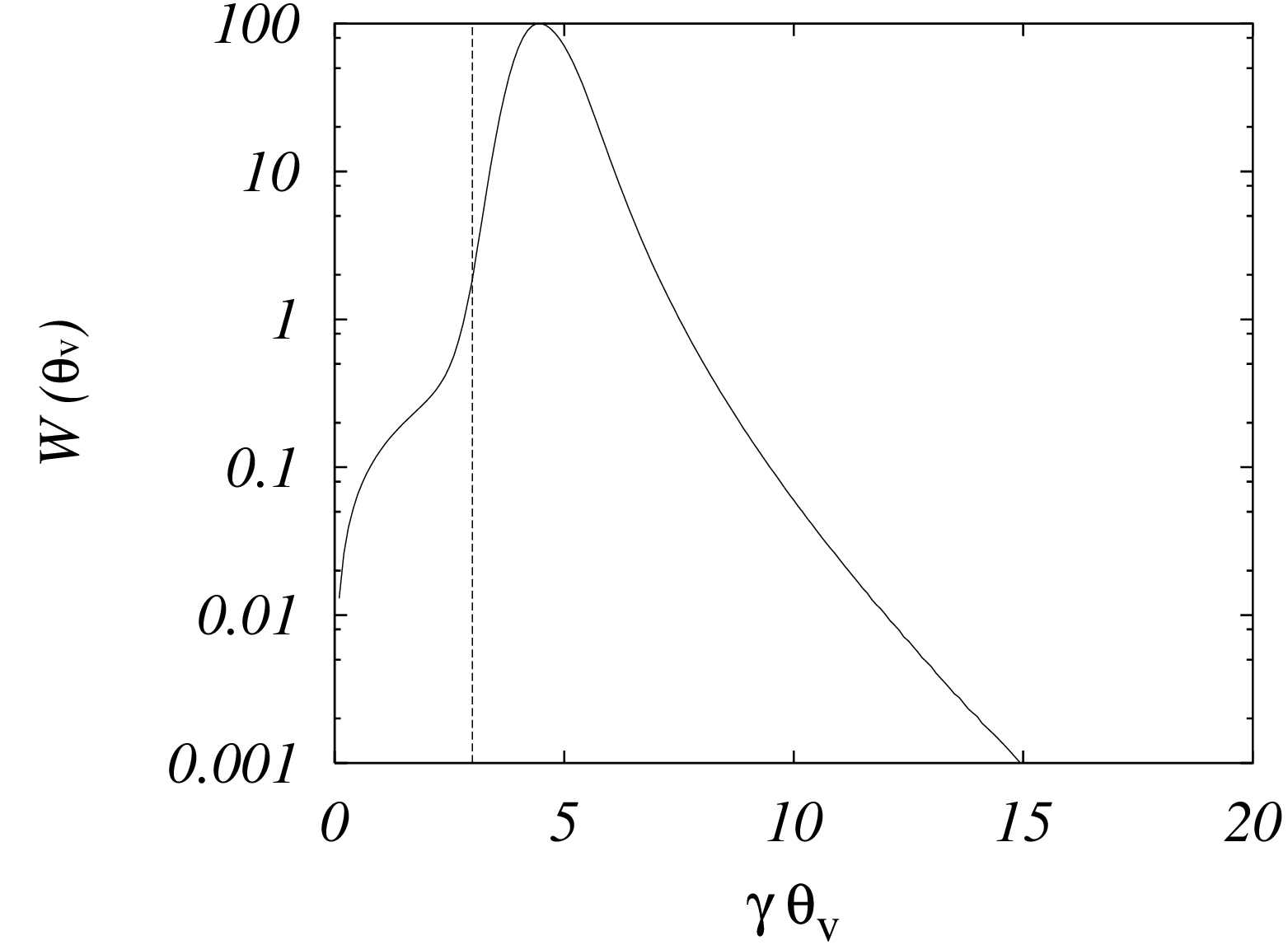}
\caption{
(Left panel):
The maximum (minimum) redshift, $z_{\rm max}$ ($z_{\rm min}$),
of the XRF as a function
of the viewing angle $\gamma\theta_v$ is shown as
the solid line (dashed line).
The jet emission is observed as the XRF if the source has a redshift
$z$ in the range $z_{min}<z<z_{max}$. \
(Midle panel):
$\VVM_{\Delta\theta,\,\theta_v}$ for the XRF detected by the 
WFCs/{\it BeppoSAX} is shown as a function of $\gamma\theta_v$. \
(Right panel):
The weight function $W(\theta_v)$ (arbitrary normalization), 
which is the relative observed event rate,
is shown as a function of $\gamma\theta_v$. \
All figures are for the case of $\Delta\theta=0.03$.
The vertical dashed lines represent $\theta_v=\Delta\theta=0.03$. 
}
\label{figs}
\end{figure}

The $\VVM$ for fixed opening half-angle $\Delta\theta$ is
calculated as
\begin{equation}
\VVM_{\Delta\theta}=
\f{\int
\VVM_{\Delta\theta,\,\theta_v}
\, W(\theta_v)\,d\theta_v}
{\int W(\theta_v)\,d\theta_v}\ ,
\label{vvm3}
\end{equation}
where $\VVM_{\Delta\theta,\,\theta_v}$ is for fixed
$\Delta\theta$ and $\theta_v$
(See \citet{yin03b} for details).
The weight function $W(\theta_v)$ is
the product of the solid angle factor and the volume factor:
\begin{equation}
W(\theta_v)=2\pi\sin\theta_v
\int_{z_{\rm min}(\theta_v)}^{z_{\rm max}(\theta_v)}dz\,
\f{n(z)}{1+z}\,4\pi \left(\f{d_L}{1+z}\right)^2\,
\f{d}{dz}\left(\f{d_L}{1+z}\right) \ ,
\end{equation}
where $n(z)$ and $d_L$ are the comoving GRB rate density and
the luminosity distance, respectively.
Here $z_{\rm max}$ ($z_{\rm min}$) is the maximum (minimum) redshift 
of the XRF for given $\Delta\theta$ and $\theta_v$.
In determining  $z_{\rm min}$  and $z_{\rm max}$, we should  note 
that the operational definition of the {\it BeppoSAX}-XRF is the fast 
X-ray transient with duration less than $\sim10^3$ seconds which is
detected by WFCs and not detected by the GRBM \citep{He01a}.
Therefore, if the sources are nearby such that $z<z_{\rm min}$,
they are observed as GRBs because the observed fluence
in the $\gamma$-ray band becomes larger than
the limiting sensitivity of GRBM 
($\sim 3\times10^{-6}{\rm ergs}\ {\rm cm}^{-2}$).
If the sources are too far such that $z>z_{\rm max}$,
they cannot be observed by WFCs with  a limiting sensitivity
of about $4\times10^{-7}{\rm ergs}\ {\rm cm}^{-2}$.
The behavior of $z_{\rm max}$\,, $z_{\rm min}$\,,
$\VVM_{\Delta\theta,\,\theta_v}$\,, and $W(\theta_v)$
for $\Delta\theta=0.03$ are shown in  Figure~\ref{figs}.
%
%

\section{DISCUSSION}
The results of the numerical integration
are summarized in Table~\ref{TableVVM}.
For each $\Delta\theta$,
$z_{\rm max}(\theta_{v,\, p})$ [$z_{\rm min}(\theta_{v,\, p})$]
means the maximum (minimum) redshift where $W(\theta_v)$ takes the
maximum value.
If we take the jet opening half-angle as 
$\Delta\theta\lesssim0.03$, $\VVM_{\Delta\theta}$ is smaller 
than $\sim0.4$, which is consistent with the observational 
result at the  1\,$\sigma$ level.
The value of $\Delta\theta\sim0.03$ is as low as
the minimum of those having ever been inferred from afterglow
light curve.
The jet break time is given by 
$t_j\sim 13$\,min\,$(\Delta\theta/0.01)^{8/3}$ \citep{fra01}
and so it requires fast localization to observe the jet break
for a narrow jet.
Therefore, at present, the small number of GRBs with small
$\Delta\theta$ may come from the observational selection effect.
In the context of this scenario, we might be able to account 
for the fact that
afterglows of XRFs have been rarely observed since the afterglow 
at a fixed time gets dimmer for an earlier break time.
Furthermore, some ``dark GRBs''
might be such a small opening angle jet observed
with an on-axis viewing angle for the same reason.

Table~\ref{TableVVM} shows that
the sources with the viewing angle 
$\theta_{v,\,p}\sim\Delta\theta+0.02$,
where $W(\theta_v)$ takes maximum,
are the most frequent class of the XRFs in the population for
$\Delta\theta\lesssim0.03$.
The typical observed photon energy is estimated as 
\citep{yin02,yin03b}
\begin{equation}
E_p\sim 2\gamma\nu'_0(1+z)^{-1}
[1+(\gamma\theta_{v,\,p}-\gamma\Delta\theta)^2]^{-1}
\sim30\ {\rm keV}\ [(1+z)/2.5]^{-1}\ ,
\end{equation}
which is the typical observed peak energy of the XRFs
\citep{ki02}.
We can propose from our argument that
the emissions from the jets with a small opening half-angle 
such as $\Delta\theta\lesssim0.03$ are 
observed as XRFs when they are seen from off-axis viewing angle.
If one can detect the afterglow of the XRF,
which has the maximum flux at about several hours after the XRF,
the fitting of light curve may give us the key information
about the jet opening angle \citep{gpkw02}.
For example, the light curve of afterglow of XRF\,030723 was unusual
in early epoch,  which can be well explained if the jet is seen from 
off-axis viewing angle \citep{huang03}.

We can estimate the observed event rate of the XRF 
for fixed $\Delta\theta$ as
$R^{{\rm XRF}}_{\Delta\theta}=(1/4\pi)\int W(\theta_v)\,d\theta_v$\,.
For a reasonable proportionality constant,
 we derive
$R^{{\rm XRF}}_{\Delta\theta=0.03}\sim 
10^2 \ {\rm events} \ {\rm yr}^{-1}$,
which is consistent with the observation.
The value of $R^{{\rm XRF}}_{\Delta\theta}$ remains 
unchanged within 
a factor of 2 when we vary $\Delta\theta$ from 0.01 to 0.07.

When the jet opening half-angle $\Delta\theta$ has a distribution
$f_{\Delta \theta}$,
we integrate $\VVM_{\Delta\theta}$ and $R^{\rm XRF}_{\Delta\theta}$ 
over the distribution of $\Delta\theta$ as
\begin{equation}
\VVM\propto
\int d(\Delta\theta)\,
f_{\Delta\theta}\,R^{\rm XRF}_{\Delta\theta}\,
\VVM_{\Delta\theta}\  ,
\end{equation}
\begin{equation}
R_{{\rm XRF}}\propto
\int d(\Delta\theta)\,f_{\Delta\theta}\,
R^{\rm XRF}_{\Delta\theta} \ ,
\end{equation}
respectively \citep{yin03b}.
When we adopt a power-low distribution as
$f_{\Delta\theta}\propto(\Delta\theta)^{-q}$, with
$q=4.54$ \citep{fra01} and 
integrate over $\Delta\theta$ from 0.01 to 0.2 rad,
we find $\VVM=0.36$ and 
$R_{{\rm XRF}}\sim10^2$~events~yr$^{-1}$.
These values mainly depend on the lower cut-off of
$f_{\Delta\theta}$.
For example, we obtain $\VVM=0.43$ and 
$R_{{\rm XRF}}\sim3$~events~yr$^{-1}$ if the integration is done 
over $\Delta\theta$ from 0.03 to 0.2 rad.
Hence, we might be able to determine the lower cut-off 
if the other uncertain factors are fixed by other arguments.
Since the statistics of the observational data will increase
in the near future owing to instruments such as
{\it HETE-2} and {\it Swift},
we will be able to say more than above discussion, 
including more accurate functional form of $f_{\Delta\theta}$
than that we have considered above,
as well as the relation to the GRB event rate.

\begin{theacknowledgments}
This work was supported in part by
Grant-in-Aid for Scientific Research 
of the Japanese Ministry of Education, Culture, Sports, Science
and Technology, No.05008 (R.Y.), No.00660 (K.I.), 
No.14047212 (T.N.), and No.14204024 (T.N.).
\end{theacknowledgments}


\IfFileExists{\jobname.bbl}{}
 {\typeout{}
  \typeout{******************************************}
  \typeout{** Please run "bibtex \jobname" to optain}
  \typeout{** the bibliography and then re-run LaTeX}
  \typeout{** twice to fix the references!}
  \typeout{******************************************}
  \typeout{}
 }

\end{document}